# Frustrated Magnetism of Pharmacosiderite Comprising Tetrahedral Clusters Arranged in the Primitive Cubic Lattice

Ryutaro Okuma[a]*, Takeshi Yajima[a], Tatsuo Fujii[b], Mikio Takano[b], and Zenji Hiroi[a]

[a]Institute for Solid State Physics, University of Tokyo, Kashiwa, Chiba 277-8581, Japan
[b]Department of Applied Chemistry, Okayama University, Tsushima-naka 3-1-1, Okayama 700-8530, Japan

We show that pharmacosiderite is a novel cluster antiferromagnet comprising frustrated regular tetrahedra made of spin-5/2 $Fe^{3+}$ ions that are arranged in the primitive cubic lattice. The connectivity of the tetrahedra and the inter-cluster interaction of 2.9 K, which is significantly large compared with the intra-cluster interaction of 10.6 K, gives a unique playground for frustration physics. An unconventional antiferromagnetic order is observed below $T_N$ ~ 6 K, which is accompanied by a weak ferromagnetic moment and a large fluctuation as evidenced by Mössbauer spectroscopy. A $\mathbf{q}$ = 0 magnetic order with the total $S$ = 0 for the tetrahedral cluster is proposed based on the irreducible representation analysis, which may explain the origin of the weak ferromagnetism and fluctuation.

Geometrically frustrated magnets with competing nearest-neighbor interactions can possess massive degeneracy around the ground state, which tends to yield exotic quantum phases such as spin liquid states.[1] In real magnets, however, perturbative interactions often lift the degeneracy and thus induce magnetically ordered states with noncollinear or noncoplanar arrangements of spins. Such peculiar orders are characterized by the emergence of higher-order degrees of freedom such as the spin chirality, which cause a variety of phenomena beyond those from the simple spin degree of freedom.[2,3]

The frustrated cluster magnet is one of frustrated magnets, which comprise clusters made of spins, instead of single spins, arranged in three-dimensional lattices; here we exclude single-molecule magnets[4] or molecular cluster magnets, such as $LiZn_2Mo_3O_8$,[5] with unpaired electrons in the molecular orbitals of the cluster. A spin cluster made of regular triangles such as a tetrahedron naturally possesses an internal degree of freedom and thus nontrivial degeneracy in the ground state. A coupling between the clusters in a frustrated cluster magnet may cause a nontrivially ordered state or a quantum-mechanically disordered state arising from the complex degrees of freedom of the cluster.

Here, we focus on frustrated cluster magnets made of regular tetrahedral clusters. The $T_d$ point group symmetry of one regular tetrahedron is globally preserved in the three types of extended structures depicted in Figs. 1(a)–(c), in which regular tetrahedra form face-centered cubic (fcc), body-centered cubic (bcc), and primitive cubic (pc) lattices, respectively. In addition to the magnetic interaction $J$ in the tetrahedron, there are inter-cluster interactions $J'$ which form a regular tetrahedron for the fcc lattice and elongated tetrahedra for the bcc and pc lattices. Thus, both intra- and inter-cluster couplings are frustrated in these unique systems.

The fcc lattice of tetrahedral clusters is called the breathing pyrochlore lattice[6] and is realized in such materials as $LiGa_{1-x}In_xCr_4O_8$[6,7] and $Ba_3Yb_2Zn_5O_{11}$.[8] They show exotic magnetism arising from the competition between $J$ and $J'$ at $J'/J$ < 0.6. A classical spin nematic state is realized in the former compound at $x$ = 0.05,[9] while the ground state of the latter compound is a unique gapped state comprising localized spin singlets.[10] The bcc lattice is realized in $Co_4B_6O_{13}$,[11] helvine,[12] and danalite,[13] and the pc lattice is realized in pharmacosiderite[14] and $K_{10}M_4Sn_4S_{17}$ ($M$ = $Mn^{2+}$, $Fe^{2+}$, $Co^{2+}$).[15] However, the details of their magnetic properties have not yet been known.

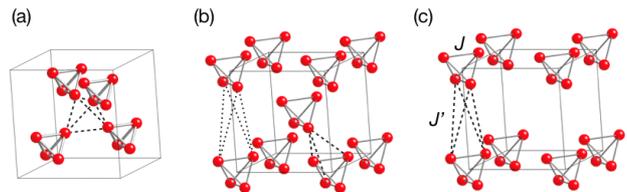

**Fig. 1**. (Color online) Three kinds of cubic lattices comprising regular tetrahedra: (a) face-centered lattice (breathing pyrochlore lattice), (b) body-centered lattice, and (c) primitive lattice. The solid and dashed lines represent intra- ($J$) and inter-cluster ($J'$) magnetic interactions, respectively.

We focus on pharmacosiderite. It crystallizes in the space group $P\bar{4}3m$ with the lattice constant of $a$ = 7.98 Å.[14] As depicted in the inset of Fig. 2, four $FeO_6$ octahedra coupled by edge sharing are located at the corner of the cubic unit cell, which contain a tetramer made of $Fe^{3+}$ ions. The tetramers are bridged through tetrahedral $AsO_4$ units to form a zeolitic framework, in which a large monovalent cation or water molecules are accommodated in the void.

On the physical properties, pharmacosiderite is a Mott insulator with the $Fe^{3+}$ ion carrying isotropic spin 5/2 in the orbital-quenched $d^5$ electron configuration. About half century ago, Takano and coworkers gave a brief report on the magnetic susceptibility and Mössbauer effect:[16] the former shows a broad maximum around 20 K, and the latter shows a broadened six-line spectrum below 6 K, from which an antiferromagnetic order is suggested.

In this letter, we report synthesis, magnetization, heat capacity, and Mössbauer effect using polycrystalline samples of pharmacosiderite. We show that pharmacosiderite is a unique frustrated cluster magnet made of strongly coupled spin tetrahedra. A **q** = 0 magnetic order with a large fluctuation is suggested.

A polycrystalline sample of pharmacosiderite has been synthesized by the conventional hydrothermal method.[17] First, a beige colloidal precursor is obtained by vigorous stirring of a ferric solution (6 g of $NH_4Fe(SO_4)_2 \cdot 12H_2O$ fully dissolved in 1.5 g of water) and an arsenate solution (1.6 g of $KH_2AsO_4$ and 1.5 g of $K_2CO_3$ fully dissolved in 3.8 g of water). After the pH of the precursor is adjusted to ca. 1.5 by the addition of 0.02 ml of 10M HClaq, it is heated in a Teflon-lined autoclave for three hours at 220 ˚C. After the reaction, a pale-yellow precipitation is filtered off and thoroughly washed with water. Then, the obtained powder is annealed in 500 ml of 0.1M HClaq for a week at 100 ˚C. By this annealing, the crystallinity has been improved with water molecules exclusively occupying the zeolitic voids. Finally, water-containing pharmacosiderite $(H_3O)Fe_4(AsO_4)_3(OH)_4 \cdot 5.5H_2O$ is obtained as a pale-green powder after filtration and drying.

Thus obtained samples were examined by powder X-ray diffraction (XRD) at temperatures down to 4 K. As shown in Fig. 2, a Rietveld refinement based on the structural model of ref. 12 with lattice constant of $a$ = 8.00560(4) Å is successfully converged to $R_{wp}$ = 7.602%, $R_p$ = 5.381%, $S$ = 1.6046 at 4 K. Neither peak splitting nor appearance of additional peaks was observed, indicating the absence of structural transitions. Chemical analysis by the X-ray spectrometry yields the atomic ratio of Fe:As = 4:3.04(3) with no trace of potassium, which is consistent with the ideal chemical formula.

Magnetization and heat capacity measurements were performed in the temperature range of 2–300 K using Quantum Design MPMS3 and PPMS, respectively. $^{57}$Fe Mössbauer spectra of a powder sample were taken in a conventional constant-acceleration spectrometer.

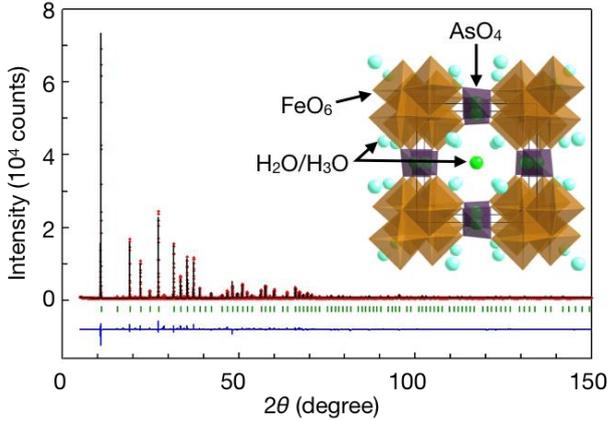

**Fig. 2**. (Color online) Powder XRD profile of pharmacosiderite at 4 K fitted by the Rietveld method, showing observed (red cross), calculated (black line), and difference (blue line). The ticks represent the positions of the Bragg reflections of pharmacosiderite. The inset shows the crystal structure containing the units of FeO$_6$ and AsO$_4$ depicted by brown octahedra and purple tetrahedra, respectively. The balls represent H$_2$O and H$_3$O$^+$ molecules.

The magnetic susceptibility shows an upturn below 6 K after a broad peak around 13 K [Fig. 3(a)]. The broad peak is similar to that reported in the previous study,[16] while the upturn is much larger. The heat capacity shows a peak around 6 K, which evidences a long-range antifferromagnetic order in bulk below $T_N$ ~ 6 K. Fitting the susceptibility at 150–300 K to the Curie–Weiss law yields the Curie–Weiss temperature $T_{CW}$ = –155.8(6) K, the effective moment $p_{eff}$ = 5.982(4), and the temperature-independent term $\chi_0$ = –5.2(3)×10$^{-5}$ cm$^3$ mol$^{-1}$. The Landé $g$ factor for $S$ = 5/2 is 2.023(2), indicating a negligible spin–orbit coupling.

The broad peak in the magnetic susceptibility is reproduced by assuming a coupled cluster model in the mean-field approximation:[18]

$$\chi(T,S,J,J') = \frac{\chi_{tetra}(T,S,J)}{T-zJ'\chi_{tetra}(T,S,J)},$$

where $\chi_{tetra}(T,S,J)$ represents the spin susceptibility of one regular tetrahedron made of spin $S$ coupled by the Heisenberg interaction $J$ at temperature $T$, $J'$ is the interaction between spins between nearby clusters, and $z$ is the number of the $J'$ bonds ($z$ = 6). As shown in Fig. 3, a fitting in the temperature range of 13–300 K looks almost perfect, giving both antiferromagnetic $J$ = 10.6 K and $J'$ = 2.9 K. Thus, pharmacosiderite is a frustrated cluster magnet with the moderately large inter-cluster coupling via the AsO$_4$ unit.

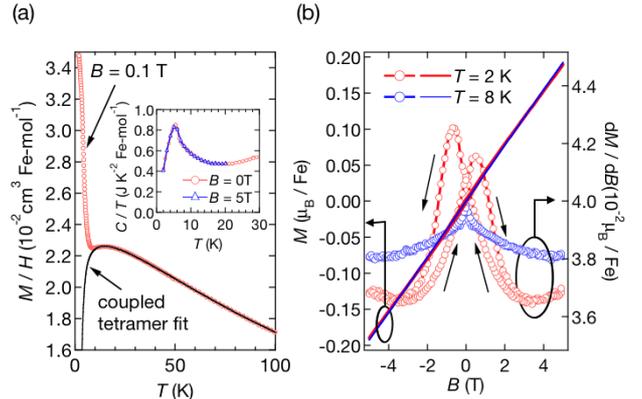

**Fig. 3**. (Color online) (a) Temperature dependence of magnetic susceptibility measured at $B$ = 0.1 T on a polycrystalline sample of pharmacosiderite. The black curve represents a fit to the coupled cluster model mentioned in the text, which gives $J$ = 10.6 K and $J'$ = 2.9 K. The inset shows the temperature dependences of heat capacity at $B$ = 0 and 5 T. (b) Magnetizations (left) and its field derivatives measured at 2 and 8 K. A small hysteresis is observed between –2 and 2 T only at 2 K.

The magnetization process at 2 K in Fig. 3(b) shows a tiny hysteresis, as is clear in its derivative, which suggests the presence of a tiny ferromagnetic moment of 6.9(6)×10$^{-3}$ μ$_B$. This net moment appears below 6 K and thus is associated with the magnetic order below $T_N$. Thus, the upturn in the temperature dependence of susceptibility is attributed to the weak ferromagnetic moment.

The temperature evolution of Mössbauer spectra is shown in Fig. 4. The spectrum at 6 K shows only a quadruple splitting indicating a paramagnetic state above $T_N$. The spectrum at 4 K becomes broadened but not magnetically split, in spite that a long-range order with a weak ferromagnetic moment has already occurred at this temperature. At 2.8 K, a six-line spectrum is eventually observed as previously reported.[16] However, even at this temperature well below $T_N$, the peaks remain broad, which indicates that the spins significantly fluctuate owing to certain reason in the ordered state, which must be associated with the frustration of and between tetramers in the pc lattice.

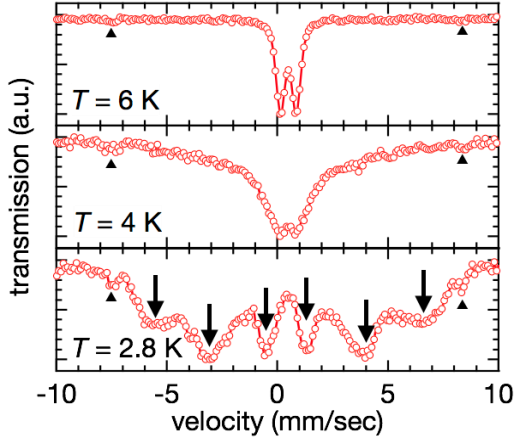

**Fig. 4.** (Color online) Mössbauer spectra of a powder sample of pharmacosiderite at $T$ = 2.8, 4, and 6 K. The black arrows indicate the six-lines spectrum in the magnetically ordered state. The black triangles show signals from tiny amount of the impurity phase of jarosite.

To summarize our main experimental results on pharmacosiderite, the cubic symmetry is preserved down to 4 K, and a magnetic order sets in below $T_N \sim 6$ K with a tiny ferromagnetic moment. Both the intra- and inter-cluster interactions are antiferromagnetic, 10.6 and 2.9 K, respectively. A large magnetic fluctuation persists well below $T_N$.

Now we discuss the magnetic structure of pharmacosiderite and the origin of the fluctuation. For an isolated cluster, the sum of four spins should be zero in the ground state. To align these clusters in the pc lattice, let us consider the simple case with each cluster possessing the identical spin configuration; that is the **q** = 0 structure. Since the transition of pharmacosiderite is probably of the second order, a possible magnetic structure should belong to one of the irreducible representations in the representational analysis; the representational analysis for the pyrochlore lattice[19-21] is applied to the present case of the pc lattice.

Taking into account the presence of the weak ferromagnetic moments, the $\Gamma_5$ irreducible representation is selected because only this includes ferromagnetic spin configurations in addition to antiferromagnetic ones. Taking the z axis unique, an allowed antiferromagnetic spin arrangement is a coplanar structure made of two pairs of antipararell spins pointing along the directions of the tetrahedron edges perpendicular to the z axis, as shown in Fig. 5(a). A linear combination of this basis vector with one of the ferromagnetic basis vectors results in a spin canting along the [001] direction. A possible source for such a spin canting is the single-ion anisotropy toward the local threefold rotational axis along the <111> direction. Note that a very large single-ion anisotropy would generate the so-called "two-in–two-out" spin structure such as observed in spin-ice compounds,[22] but this is not the case for pharmacosiderite with weak anisotropy that gives a small canting. This type of order is consistent with our recent magnetization measurements on a single crystal of pharmacosiderite, which will be reported elsewhere.

On the three-dimensional arrangements of the spin clusters, other propagation vectors such as (1/2, 1/2, 1/2) may not occur, because nearly ferromagnetic inter-cluster couplings are required for them. Incommensurate structures cannot be excluded but may be unlikely for the simple $J$–$J'$ model on the pc lattice.

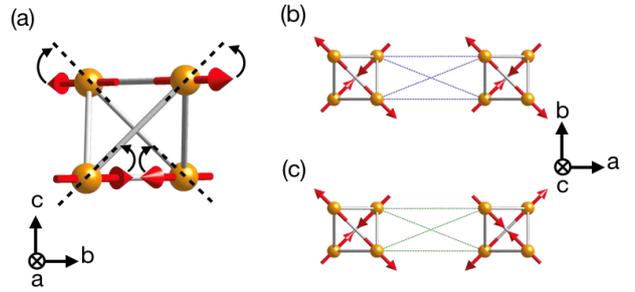

**Fig. 5.** (Color online) (a) Proposed spin configuration for one $Fe^{3+}$ tetrahedron in pharmacosiderite. Two pairs of antiferromagnetically-coupled spins are aligned along the [110] and [1–10] direction. All the spins can be slightly inclined toward the local [111] easy axes depicted by the dashed lines, which produces a net ferromagnetic moment along [001]. A pair of tetrahedra viewed along [001] is shown for **q** = 0 (b) and (1/2, 1/2, 1/2) (c), in which antiferromagnetic and ferromagnetic inter-cluster coupling $J'$s occur, respectively.

The reason why the $\Gamma_5$ magnetic structure with **q** = 0 is selected from many degenerate states in pharmacosiderite is likely the Dzyaloshinskii–Moriya (DM) interaction which is usually a leading anisotropy term in orbitally nondegenerate system.[23] In fact, this is the case for a pyrochlore lattice antiferromagnet with one of two symmetry-allowed sets of DM vectors.[24] The same argument may apply to the pharmacosiderite.

Finally, we comment on the origin of the fluctuation below $T_N$. The observed broad Mössbauer spectra indicate that spins can fluctuate even in the ordered state, which is unusual for a classical spin system. The Mössbauer spectrum at 2.8 K is not reproduced by the Blume-Tjon model, which assumes a randomly fluctuating internal magnetic field of the Ising type.[25] Probably, a certain correlated fluctuation is to be considered.

We think that the origin is related to the unique frustration of the $\Gamma_5$ magnetic structure with **q** = 0 in the pc lattice: there must occur magnetic domain walls without energy loss or zero-energy excitation modes, which may partially destroy the long-range order and cause magnetic fluctuations, the detail of which will be discussed elsewhere. Neutron scattering experiments are now in progress to unveil the magnetic structure and intriguing spin dynamics of pharmacosiderite.

**Acknowledgments** We are grateful to Tsuyoshi Okubo and Nic Shannon for helpful discussion. RO is supported by the Materials Education Program for the Future Leaders in Research, Industry, and Technology (MERIT) given by the Ministry of Education, Culture, Sports, Science and Technology of Japan. This work was partly supported by the Core-to-Core Program for Advanced Research Networks given by the Japan Society for the Promotion of Science (JSPS).